# Tension-oriented cell divisions limit anisotropic tissue tension in epithelial spreading during zebrafish epiboly

Pedro Campinho[1], Martin Behrndt[1], Jonas Ranft[2,3,4,5], Thomas Risler[2,3,4], Nicolas Minc[6] and Carl-Philipp Heisenberg[1,*]

[1] Institute of Science and Technology Austria, Am Campus 1, A-3400 Klosterneuburg, Austria
[2] Institut Curie, Centre de Recherche, F-75005 Paris, France
[3] UPMC Univ Paris 06, UMR 168, F-75005 Paris, France
[4] CNRS, UMR 168, F-75005 Paris, France
[5] Max-Planck-Institute for the Physics of Complex Systems, Nöthnitzer Strasse 38, D-01187 Dresden, Germany
[6] Institut Jacques Monod, CNRS, UMR 7592, 15 rue Hélène Brion, 75205 Paris , France
* Corresponding author (heisenberg@ist.ac.at)

**Abstract**

**Epithelial spreading is a common and fundamental aspect of various developmental and disease-related processes such as epithelial closure and wound healing. A key challenge for epithelial tissues undergoing spreading is to increase their surface area without disrupting epithelial integrity. Here we show that orienting cell divisions by tension constitutes an efficient mechanism by which the Enveloping Cell Layer (EVL) releases anisotropic tension while undergoing spreading during zebrafish epiboly. The control of EVL cell-division orientation by tension involves cell elongation and requires myosin II activity to align the mitotic spindle with the main tension axis. We also found that in the absence of tension-oriented cell divisions and in the presence of increased tissue tension, EVL cells undergo ectopic fusions, suggesting that the reduction of tension anisotropy by oriented cell divisions is required to prevent EVL cells from fusing. We conclude that cell-division orientation by tension constitutes a key mechanism for limiting tension anisotropy and thus promoting tissue spreading during EVL epiboly.**

**Introduction**

In zebrafish gastrulation, the EVL is formed as a squamous epithelial cell layer at the animal pole of the embryo and spreads over the entire spherical yolk cell during the course of epiboly (4–10 hours post fertilization (hpf)), thereby rapidly increasing its surface area[1]. Actomyosin contraction within the yolk syncytial layer (YSL), to which the EVL is connected at its margin, is thought to drive EVL epiboly movements by pulling on the EVL margin in direction of the vegetal pole[2-4]. However, the mechanisms by which the epibolizing EVL rapidly increases its surface area and, at the same time, maintains its epithelial integrity during zebrafish epiboly remain unclear.

Studies in another teleost fish, *Fundulus heteroclitus*, have suggested that the EVL increases its surface area during epiboly by both passive cell spreading in response to pulling forces from the YSL, and active cellular rearrangement adjusting the shape of the EVL to the spherical geometry of the yolk cell on which it spreads[5]. Consistent with the idea of EVL cells passively spreading as a result of the YSL pulling on the EVL margin are observations in zebrafish that marginal EVL cells become increasingly elongated along the axis of tissue spreading and in *Fundulus* of tension building up within the plane of the EVL during the course of epiboly[4,5]. Cellular rearrangements have been noted to particularly occur at the margin of the EVL in both zebrafish and *Fundulus* with individual cells constricting at their leading edge and eventually being displaced from the EVL margin[4,6]. However, whether EVL cell spreading and rearrangement constitute the sole mechanisms mediating EVL surface expansion during fish epiboly, or whether other mechanisms might also be involved, is still unclear.

Cell divisions have profound effects on epithelial tissue morphogenesis[7], and it is thus conceivable that they also play a critical role in EVL epiboly movements. In zebrafish, EVL cells have been shown to undergo divisions within the plane of the tissue during epiboly[8]. Moreover, planar tension within the EVL has been speculated to function in maintaining EVL cell lineage by keeping the mitotic spindle of dividing cells oriented in the plane of the cell sheet[8,9]. Here, we investigate how cell divisions contribute to EVL epiboly movements, and how tension within the EVL relates to the EVL cell-division rate and/or orientation.

**Results**

**Anisotropic tissue tension controls cell division orientation within the EVL**

To obtain insight into the cellular processes underlying EVL tissue spreading during zebrafish epiboly, we first asked how the increase of total surface area of the EVL during epiboly correlates with changes in surface area, height and volume of individual EVL cells and with their division pattern (Fig. 1 and Supplementary Video 1). We found that EVL tissue spreading is accompanied by pronounced EVL cell flattening, as recognized by an increase in apical surface area and concomitant decrease in cell height along the apical-basal axis of individual EVL cells (Fig. 1b and Supplementary Fig. 1a,b). In contrast, EVL cell volume remained largely constant during flattening and was cut in half once these cells underwent division (Fig. 1c and Supplementary Fig. 1c). EVL cells exclusively divided within the plane of the epithelium with higher frequency at early stages of EVL epiboly and at central/animal compared to marginal regions of the EVL (Fig. 1d). Moreover, our observation that the total number of EVL cells more than doubles during the early stages of epiboly (Fig. 1e) suggests that every EVL cell on average divides at least once within this period. Taken together, these observations suggest that the EVL tissue spreads by cell flattening without significant increase in tissue volume.

Oriented cell divisions have been hypothesized to facilitate tissue spreading[10]. To investigate their role in EVL spreading, we first analyzed division orientation of individual EVL cells during the course of epiboly. We found that EVL cells preferentially divided along the axis of EVL tissue spreading (animal-vegetal axis of the embryo) during all stages of epiboly (Fig. 1f and Supplementary Video 1). This stereotypical division orientation (SDO) of EVL cells is consistent with a potential role of oriented cell divisions in EVL epiboly progression.

We next asked how SDO of EVL cells is controlled during EVL epiboly. Mechanical tension has been suggested to represent a powerful mechanism for orienting the divisions of isolated cells in culture[11]. Moreover, studies in *Fundulus heteroclitus* have suggested that the epibolizing EVL is under tension both around the circumference and along the animal-vegetal axis of the embryo[5,12]. We therefore hypothesized that anisotropic tension distribution within the EVL may control SDO of EVL cells. To test this hypothesis, we used a laser-cutting device to map tension within the EVL during the course of epiboly[2,13]. To detect tension along the animal-vegetal axis of the EVL, we cut the apical actomyosin cortex of the EVL along a 100-µm-long line parallel to the EVL margin and determined the recoil velocity of the cortex as a readout for animal-vegetal tension (Fig. 2a). Similarly, to detect EVL tissue tension perpendicular to the animal-vegetal axis, we determined the recoil velocity of the actomyosin cortex in cuts oriented perpendicular to the EVL margin (Fig. 2a and Supplementary Video 2). We found that tension along the animal-vegetal axis of the EVL is higher than along the circumference of the embryo, and that this anisotropic tension distribution is particularly pronounced at mid- to late-gastrulation stages (Fig. 2b). This global correlation between EVL tissue tension anisotropy and SDO of EVL cells is consistent with a function of tissue tension in orienting EVL cell divisions.

To determine whether there is a causal link between EVL tissue tension and cell-division orientation, we locally induced anisotropic tissue tension within the EVL and analyzed resultant changes in cell-division orientation. We locally induced anisotropic tension within the EVL by simultaneously ablating two small groups of EVL cells (2–3 cells each) positioned close to each other ($\approx$ 120 µm), which led to the extrusion of the ablated cells and, consequently, considerable stretching of the EVL tissue between the two ablation sites (Fig. 2c and Supplementary Video 3). We then analyzed changes in spindle orientation of cells that were located between the two ablation-sites and had their metaphase spindle axis oriented perpendicular to the axis of induced tension before tension was applied (Fig. 2c). We found that in these cells the mitotic spindle preferentially re-oriented along the main axis of tension before cytokinesis (Fig. 2c), demonstrating that tissue tension anisotropy influences the cell-division orientation of EVL cells.

**The orientation of EVL cell divisions by tension involves cell elongation and requires myosin II activity**

We next asked how anisotropic tissue tension controls EVL cell-division orientation. Many cell types have been shown to preferentially orient their mitotic spindle along their longest axis[14-17]. Notably, EVL cells display flat two-dimensional defined geometries and do not round up at their apical side during mitosis (Supplementary Video 4), and thus cell-shape anisotropies could influence metaphase spindle orientation through putative length-dependent astral microtubule forces[14]. Considering that EVL cells appear preferentially elongated along the main axis of tension shortly before undergoing cytokinesis (animal-vegetal axis; Fig. 3a), we therefore hypothesized that anisotropic EVL tissue tension may control SDO of EVL cells by elongating cells along the main axis of tension. To assess the potential role of EVL cell elongation on division orientation, we made use of a computational model that predicts the preferred division orientation based on cell shape[14]. We found that EVL cell elongation is a reliable predictor for division orientation (Figs. 3b,c), supporting our hypothesis that EVL tissue tension controls cell-division orientation by cell elongation.

In addition to cell shape, external forces have also been proposed to control spindle orientation in a myosin II-dependent manner[11]. To determine whether myosin II is required for tension-controlled EVL cell-division orientation, we analyzed division orientation in embryos with normal and reduced myosin II activity. To reduce myosin II activity, we exposed embryos to the myosin II inhibitor Blebbistatin. We then compared the effect of cell elongation on division orientation between control and Blebbistatin-treated cells and found that EVL cell shape less accurately predicted spindle orientation in cells with reduced myosin II activity (Fig. 3b,c). Moreover, the failure of the mitotic spindle to align with the longest axis of the cell and thus the main axis of tension in Blebbistatin-treated embryos was often accompanied by increased spindle fluctuations (Fig. 3d and Supplementary Video 5). The effect of myosin II inhibition on spindle orientation could in principle be due to defects in cell elongation and/or spindle alignment with the longest cell axis. To distinguish between these two possibilities, we compared EVL cell elongation between control and Blebbistatin-treated embryos and analyzed how accurately our computational model can predict experimental spindle orientation on the basis of these measured cell shapes. Consistent with a critical function of actomyosin contraction in cell shape changes[18], we found that EVL cell elongation was reduced in Blebbistatin-treated embryos (Fig. 3e), suggesting that myosin II affects spindle orientation by cell elongation. However, even when taking these differences in cell shape into account, myosin II inhibition still increased the deviation of experimental spindle orientation in EVL cells from model predictions made on the basis of cell shape only (Fig. 3f). This comparison of model predictions with experimental observations suggests that myosin II, in addition to controlling EVL cell elongation, might be required to align the mitotic spindle with the longest cell axis.

**Tension-oriented EVL cell divisions can reduce tissue tension anisotropy and facilitate tissue spreading**

Having shown that anisotropic tissue tension can reorient cell divisions along the main axis of tension, questions arise as to the function of these tension-oriented cell divisions for EVL epiboly movements. It has been suggested that oriented cell divisions facilitate tissue spreading by decreasing tissue tension along the axis of division[10]. We therefore first asked whether individual cell divisions affect EVL tissue tension. On the characteristic timescale of a single cell-division event, stresses due to individual cell deformations may have relaxed but deformations of the cellular junctional network imply persistent elastic stresses. When modeling the forces exerted by a dividing cell as a point force dipole in a continuous elastic medium, tensions and elastic deformations show that a cell division decreases elastic stresses in the surrounding tissue along the axis of division (Fig. 5a and Supplementary Note). To test this prediction for the EVL, we turned to our minimal tension assay where we locally induce anisotropic tension within the EVL by ablating two small groups of EVL cells in close distance from each other (Fig. 4a, left). We first compared the degree of EVL tension anisotropy in cases where a cell division oriented along the axis of induced tension occurred between the ablation sites, to cases where no such cell division was observed. We found that tension anisotropy was significantly diminished in the presence of tension-oriented cell divisions (Fig. 4a, right and Supplementary Video 6), indicating that EVL cell divisions reduce tissue tension along their axis of division. We then determined whether the reduction of tissue tension along the axis of cell division also facilitates EVL tissue spreading by comparing the degree of tissue spreading in cases where a tension-oriented cell division occurred between the ablation sites, to cases without such division. Theoretically, the displacement field following a cell division shows maximum displacement along the division axis (Supplementary Note). To monitor tissue spreading in our ectopic tension assay, we analyzed the degree by which the EVL extends along the ectopic tension axis between the two ablation-sites (Fig. 4b, left). We found that EVL spreading was significantly increased in the presence of tension-oriented cell divisions (Fig. 4b, right and Supplementary Video 7), indicating that the reduction of tissue tension along the axis of cell division facilitates tissue spreading within the EVL.

**Tension-oriented EVL cell divisions prevent ectopic cell fusions**

To investigate the consequences of these findings made at the level of individual or small groups of cells for overall EVL spreading, we developed a model of the EVL that is based on our observations described so far, namely that external tensions can bias the orientation of the EVL cell-division axis, and that cell divisions release tissue tension and facilitate tissue spreading along the axis of division. We describe the EVL tissue as a continuous material on length scales large compared to that of individual cells, with a rheology that incorporates the effect of tension-oriented cell divisions in an effective shear viscosity (Fig. 5a and Supplementary Note). With these ingredients, our model predicts an anisotropic tension profile along the animal-vegetal axis of the EVL and, as a result of this, a global pattern of cell-division orientation within the EVL (Supplementary Fig. 2). This prediction fits with the experimental observations reported above (Figs 2b and 1f, respectively). Furthermore, our model predicts a decrease in tissue flow and an increase in tissue tension anisotropy when the mechanism of tension-oriented cell divisions is defective (Fig. 5b, Supplementary Note and Supplementary Fig. 2). To address these predictions, we analyzed EVL epiboly progression in embryos, in which we either reduced cell divisions during gastrulation by exposing them to the cell-division inhibitors Aphidicolin and Hydroxyurea[19], or interfered with cell-division orientation by injecting them with an $\alpha$-dynein antibody[20] (Supplementary Fig. 3a). Consistent with the model predictions, embryos with strongly reduced cell divisions showed slightly reduced EVL epiboly movements at late stages of epiboly (Fig. 5c). However, EVL tissue tension and flow appeared largely normal in embryos with no preferred cell-division orientation (Fig. 5d and Supplementary Fig. 3b), contrary to our model assumptions that orienting cell divisions by tension constitutes the main adaptive mechanism for limiting tension anisotropy within the spreading EVL. Instead, we found EVL cells fusing within the plane of the epithelium when cell divisions were reduced or misoriented (Fig. 6a,b and Supplementary Video 8). These fusions occurred with no preferred orientation, and fused cells remained integrated within the EVL and expanded their apical area at similar rates as their non-fused counterparts (Fig. 6c,d). Moreover, cell fusions were not the result of incomplete cell divisions, as fusions usually did not occur between sister cells (Supplementary Fig. 4). Importantly, EVL cell fusions were accompanied by a rapid extension of the collapsing junction directly prior to the fusion (Fig. 6e), indicative of tension release along this junction. These observations led us to hypothesize that EVL cell fusions might be caused by augmented tissue tension in embryos with reduced or misoriented cell divisions, and that they might lead to a reduction of tissue tension anisotropy, partially compensating for the lack of tension-oriented cell divisions in this process. To experimentally address this hypothesis, we sought to subject the EVL to augmented tissue tension by deforming the embryos into a cylindrical shape[2] and thereby expanding their surface area, assuming that the embryo volume is conserved (Fig. 7a). Consistent with our hypothesis of augmented tissue tension inducing cell fusions, we found EVL cell fusions to be strongly increased upon surface expansion (Fig. 7b,c and Supplementary Video 9). Moreover, our previous observation that EVL epiboly movements appear largely unaffected in cylindrical embryos[2] is compatible with the assumption that cell fusions, similar to tension-oriented cell divisions, promote EVL epiboly movements by releasing ectopic tissue tension.

Yet, questions remain as to the mechanisms by which tissue tension induces EVL cell fusions and, vice versa, EVL cell fusions release tissue tension. While difficulties in predicting and/or inducing fusions of EVL cells did not allow us to directly address the reciprocal relationship between tension and cell fusion using our ectopic tension assay, our findings so far support the notion that globally elevated tissue tension can induce cell fusions, and that cell fusions can locally release tissue tension. Whether cell fusions can also reduce global tissue tension anisotropy is not yet entirely clear. However, considering that collapsing borders oriented along the main axis of tissue tension would be expected to release more tension than collapsing borders with other orientations, the combined effect of all EVL cell fusions without preferred orientation will likely reduce overall EVL tissue tension anisotropy.

**Discussion**

Our study identifies the orientation of cell division by external tension as an efficient mechanism to release anisotropic tissue tension during EVL spreading. The degree of anisotropic tissue tension thus depends on the ability of oriented cell divisions to reduce tension anisotropy and, vice versa, the stereotypical orientation of cell divisions depends on the degree of anisotropic tension within the tissue. We expect that these reciprocal dependencies balance each other so that the EVL tissue generally displays a low degree of tension anisotropy, facilitating its spreading. Consistent with this, we observe low tension anisotropy at early stages of EVL spreading when the rate of cell divisions is high, and high tension anisotropy at later stages of epiboly when the rate of cell divisions is comparatively low (Figs 1d and 2b).

Tension has been proposed to control cell-division orientation through different mechanisms. One prime effector mechanism is cell elongation, as cell-shape anisotropies are thought to influence metaphase spindle orientation through putative length-dependent astral microtubule forces[14]. Our finding that within the epibolizing EVL, spindle orientation, cell elongation and the main axis of tissue tension are all aligned to each other, suggests that tension controls EVL cell division orientation by cell elongation. However, tension has also been shown to orient the mitotic spindle in a myosin II-dependent manner independently of its effect on cell shape[11]. While we have no direct evidence for a tension mediated polarization of myosin II to orient the division plane within the EVL, our observation that myosin II might be required for spindle orientation in EVL cells in addition to its function in cell elongation, points to similar mechanisms for cell division orientation within the EVL. Yet, to prove such mechanism, further experiments will be needed to show how tension modulates myosin II activity and whether myosin II indeed mediates the function of tension in orienting the mitotic spindle in EVL cells.

Whereas there is increasing evidence for tissue tension controlling cell division orientation[16,21,22], considerably less is known about the effect of tension-oriented cell divisions on tissue tension anisotropy. Importantly, the effect of oriented cell divisions on tension anisotropy we observed is not a consequence of tissue growth along the division axis, as EVL cells do not increase their volume between divisions (Fig. 1c). Rather, tension-oriented cell divisions lead to cellular rearrangements releasing tension along this axis of division. Similarly, cell fusions are expected to release tissue tension, and thereby partially compensate for the lack of oriented cell divisions, by fusion-mediated cell rearrangements. However, as the degrees of cell rearrangements associated with division and fusion seem similar (Supplementary Videos 4 and 8) but the number of cell fusions compared to divisions is small (21 fusions versus 576 divisions per embryo from sphere to 55 % epiboly stage; for details on how the total number of fusions and divisions per embryo was estimated see Supplementary Methods), cell fusions are unlikely to be sufficient to fully compensate for the function of oriented cell divisions in reducing tissue tension anisotropy. Observations in different teleost fish species of EVL cells undergoing junctional remodelling during epiboly[4,6,23], and in flies of anisotropic tension distribution within the spreading epidermis during dorsal closure being associated with cell elongation along the axis of main tension[24], point at junctional remodelling and cell elongation as additional mechanisms for tension release within the EVL. In agreement, we observe preferential junction disassembly within the EVL along the circumference of the embryo, which appears more pronounced when tension-oriented cell divisions are impaired (Supplementary Fig. 5). Thus, anisotropic tension release during EVL tissue spreading in epiboly relies on distinct cellular mechanisms that act in concert to limit tissue tension anisotropy and facilitate epiboly movements.

## Methods and supplementary information

Methods, supplementary information and any associated references are available in the published online version of the paper at the following url:

www.nature.com/doifinder/10.1038/ncb2869

## Acknowledgements

We are grateful to Benoit Aiguy, Suzanne Eaton, Stephan Grill, Robert Hauschild and Michael Sixt for advice, and the imaging and zebrafish facilities of the IST Austria and MPI-CBG for continuous help. We are particularly thankful to Jean-Francois Joanny for discussions regarding the theory part of this work, and Buzz Baum for sharing data prior to publication. This work was supported by the IST Austria, MPI-CBG and a grant from the Fonds zur Förderung der wissenschaftlichen Forschung (FWF) (I930-B20) to C.-P.H.

## Author contributions

P.C., M.B., J.R., T.R. and C-P.H. synergistically and equally developed the presented ideas and the experimental and theoretical approaches. P.C. performed the experiments; P.C. and M.B. did the data analysis; J.R. and T.R. developed the theory;

M.B. contributed to the experimental work; N.M. contributed to the data analysis and interpretation.

**Figure legends**

**Figure-1: EVL cell-shape changes and divisions. (a)** Exemplary images of the EVL in an embryo expressing GPI-RFP to outline EVL cells (upper panel) and respective segmented cell boundaries (lower panel) with cell divisions marked in yellow; $t = 0$ min corresponds to sphere stage (4 hpf). **(b)** Apical cell area and apical-basal cell height of individual EVL cells as a function of time after sphere stage (4 hpf); plotted values, mean ± s.e.m. normalized to the average values at $t = 120$ min (area$_{120min}$ = 838 ± 9 µm$^2$, height$_{120min}$ = 9 ± 0.29 µm); $n$, number of cells/time-point (for details see Supplementary Table 1). **(c)** Volume of individual EVL cells both before and after cell division as a function of time after sphere stage (4 hpf). Vertical lines (dashed) indicate cell divisions; plotted values, mean ± s.e.m. normalized to the average value at $t = 120$ min (volume$_{120min}$ = 6536 ± 600 µm$^3$); $n$, number of cells/time-point (for details see Supplementary Table 1). **(d)** Average percentage of EVL cells undergoing divisions as a function of their position along the animal-vegetal axis for sequential stages of early epiboly; note that EVL cell divisions

become very rare after 55 % epiboly stage[25] (for 60 % epiboly stage < 0.26 % of all EVL cells, $n$ = 10 embryos), and have therefore not been spatiotemporally analysed. For sphere and 30 % epiboly stage embryos (4 and 4.66 hpf, respectively), the EVL was subdivided along its animal-vegetal axis into three bins (I, II and III), whereas for 55 % epiboly stage embryos (5.66 hpf), the EVL was subdivided into five bins (I, II, III, IV and V); plotted values, mean ± s.e.m. normalized to the number of EVL cells per bin; see **e** for number of analyzed cells and embryos. **(e)** The total number of EVL cells per stage was on average 496, 660 and 1072 for sphere (4 hpf; $n$ = 12 embryos), 30 % (4.66 hpf; n = 10 embryos), and 55 % (5.66 hpf; $n$ = 10 embryos) epiboly stage, respectively; plotted values, mean ± s.e.m. (calculated by using embryo numbers). **(f)** Rose diagram of the cell-division axes at cytokinesis (yellow; $n$ = 524 divisions, 6 embryos) for EVL cells dividing during the course of gastrulation; $P$ (division orientation) = 0.0067 (calculated by using division numbers). A, animal; V, vegetal. Number of independent experiments: 6 (area) and 5 (height) **(b)**, 5 **(c)**, 1 **(d,e)**, 6 **(f)**.

**Figure-2: EVL tissue tension and cell-division orientation. (a)** Exemplary images of embryos at 65 % epiboly stage (7 hpf) for ultraviolet laser cuts of the apical actomyosin cortex at the animal pole (A, blue) or perpendicular (red) and parallel (green) to the EVL margin in *Tg(actb2:myl12.1-eGFP)* embryos. V, vegetal. Cuts were 100 μm long and placed 3 to 6 cell rows away from the EVL margin (for parallel and perpendicular cuts); scale bar, 20 μm. **(b)** Average initial recoil velocities for ultraviolet laser cuts throughout the course of epiboly; error bars, s.e.m.; $n$, number of cuts; note that only one cut/embryo was performed; $P$ for perpendicular vs. parallel recoil velocities (40–50 %, 5–5.25 hpf) = 0.97; $P$ (50–60 %, 5.25–6.5 hpf) = 0.11; $P$ (60–70 %, 6.5–7.5 hpf) < 0.0001; $P$ (70–80 %, 7.5–8.5 hpf) = 0.17. **(c)** Alignment of the cell-division axis with the axis of induced tension in *Tg(actb2:myl12.1-mCherry)* embryo at 40 % epiboly stage (5 hpf) injected with *tau-GFP m*RNA to mark spindle microtubules. Tension was induced orthogonally to the initial axis of the spindle (yellow) by creating two constricting wounds in the EVL

(red crosses). The resulting spindle alignment (spindle axis) was determined by measuring the angle between the final spindle axis directly prior to cytokinesis and the induced tension axis. For controls, no wounds were induced and the endogenous rotation of the spindle from its initial axis was quantified in the same manner, i.e. with the initial spindle axis orthogonal to the control axis (Supplementary Methods). Histograms show the frequency distributions of the observed angles for both control and induced tension cases; $P < 0.0001$; $n$, number of divisions; note that only one division/embryo was analysed; cell contour, white; scale bar, 20 μm. Number of independent experiments = 29 **(b)**, 14 **(c)**.

**Figure-3: Effects of cell shape and myosin II on spindle orientation and positioning. (a)** Rose diagram of the orientations of the longest cell axis of EVL cells undergoing division during the course of EVL epiboly ($n$ = 514 cells, 6 embryos) 5 min before the onset of cytokinesis; $P$ (longest axis) = 0.0369 (calculated by using cell numbers). A, animal; V, vegetal. **(b)** Alignment of the observed axis of cell-division orientation determined by the spindle axis (yellow) and the predicted axis of cell division (blue) given by cell shape in dividing EVL cells of *Tg(actb2:myl12.1-mCherry)* embryos between 30–50 % epiboly stage (4.66–5.25 hpf). Embryos were injected with *tau-mCherry m*RNA to mark spindle microtubules and treated with either the myosin II inhibitor blebbistatin or its inactive enantiomer (control); cell contour, white; scale bar, 20 μm. **(c)** Histograms show the frequency distributions of angles between predicted and observed spindle axis (Supplementary Methods) for both control (17 embryos) and blebbistatin-treated embryos (16 embryos); $P$ = 0.0056; $n$ = number of divisions. **(d)** Average maximum distance observed between experimentally determined spindle center and the prediction by the shape model in myosin II-inhibitor (blebbistatin)-treated (squares, 16 embryos) and control embryos (circles, 17 embryos); error bars, s.e.m.; $P$ = 0.0330; $n$ = number of divisions. **(e)** Minimum to maximum box-and-whisker plots of cell-shape anisotropy values (arbitrary units) computed using the shape model for EVL cells in myosin II-inhibitor (blebbistatin)-treated (right, $n$ = 32 divisions, 16 embryos) and control

embryos (left, *n* = 36 divisions, 17 embryos); *P* = 0.0163. Elongated cells have a higher value of shape anisotropy than rounder cells. **(f)** Energy penalty values (arbitrary units) for individual EVL cells computed using the shape model in myosin II-inhibitor (blebbistatin)-treated (blue squares, *n* = 32 divisions, 16 embryos) and control embryos (red circles, *n* = 36 divisions, 17 embryos) plotted over the angles between predicted and observed spindle axis. The energy penalty quantifies the deviation between the observed angle of the spindle axis from that predicted by the shape model taking differences in shape anisotropy between cells into account. For instance, the same angular deviation between observed and predicted spindle orientation would result in a lower energy penalty for cells with a small degree of shape anisotropy compared to cells with a high degree of anisotropy. Thus, if the deviation between the observed and predicted spindle orientations in blebbistatin-treated embryos was due only to a lower degree of shape anisotropy in those cells **(e)**, we would expect the energy penalty to level off with similar maximal values in control and Blebbistatin-treated conditions. Instead, the energy penalties in these two conditions are significantly different (*P* = 0.0396), suggesting that the effects of Blebbistatin treatment on spindle orientation are not solely due to changes in cell-shape anisotropy. a.u., arbitrary units. Number of independent experiments: 6 **(a)**, 13 **(b-f)**.

**Figure-4: Oriented cell divisions and EVL tissue tension. (a)** Exemplary images and average initial recoil velocities for ultraviolet laser cuts of the apical actomyosin cortex perpendicular (blue) or parallel (orange) to the division axis in *Tg(actb2:myl12.1-eGFP)* embryos at 30–40 % epiboly stage (4.66–5 hpf) after induction of tension either in the presence (bottom left; bar plot – right) or absence (top left; bar plot – left) of an EVL cell division (white cell contour) oriented along the axis of tension. Cuts (blue or orange lines) were 50 μm long; error bars, s.e.m.; *n*, number of cuts; note that only one cut/embryo was performed; *P* perpendicular recoil velocities (no division) versus (division) = 0.0112; *P* perpendicular versus parallel recoil velocities = 0.0028 (no division) and = 0.7789 (division); red crosses

mark the ablation sites where wounds were induced; scale bar, 20 μm. **(b)** Exemplary images of the spreading displacement of an EVL cell (white cell contour) and average spreading displacement in *Tg(actb2:myl12.1-eGFP)* embryos at 30–40 % epiboly stage (4.66–5 hpf) after induction of tension either in the presence (bottom left; scatter plot – squares) or absence (top left; scatter plot – circles) of a cell division (white cell contour) oriented along the axis of tension. The spreading displacement corresponds to the change in distance between the edges of the ablated cells (green lines) before ablation compared to 252 s after ablation; error bars, s.e.m.; *n*, number of embryos; note that only one measurement per embryo was performed; *P* = 0.0165; cell and wound contours, blue; red crosses mark the ablation sites where wounds were induced; scale bar, 20 μm. Number of independent experiments = 16 **(a)**, 8 **(b)**.

**Figure-5: Oriented cell divisions and EVL epiboly progression. (a)** Illustration of the main ingredients of the theoretical model proposed to describe the role of oriented cell divisions for EVL tension and spreading during epiboly (Supplementary Note). A single cell division redistributes the stresses in the tissue on the timescale of the division (left) and is itself oriented by local stresses (middle). At the level of the whole tissue, different stress-relaxation mechanisms lead to an effective viscous behavior, to which tension-oriented cell divisions contribute (right). **(b)** Schematic representation of the model predictions for tissue tensions and flow during epiboly. A reduction of tension-oriented cell divisions would be expected to increase shear viscosity, which in turn would lead to increased anisotropic tensions and a reduced tissue flow. **(c)** Epiboly progression in embryos where cell division was either blocked by incubating them in cell-division inhibitors or cell-division orientation randomized by injecting them with α-dynein antibodies; control embryos for the inhibitor and dynein antibody experiments were incubated in dimethylsulphoxide or injected with the antibody supernatant/ascites, respectively; plotted values are mean epiboly percentage ± s.e.m.; *n*, number of embryos; scale bar, 100 μm. **(d)** Initial recoil velocities for ultraviolet laser cuts

perpendicular (red) and parallel (green) to the EVL margin for control and dynein antibody-injected embryos at 50–60 % epiboly stage (5.25–6.5 hpf); error bars, s.e.m.; *n*, number of cuts; note that only one cut per embryo was performed; *P* (control versus α-dynein antibody-injected embryos, perpendicular cuts) = 0.19; *P* (control versus α-dynein antibody-injected embryos, parallel cuts) = 0.51. WT, wild type. Number of independent experiments: 3 **(c)**, 5 **(d)**.

**Figure-6: EVL cell fusions in embryos with reduced or misoriented cell divisions. (a)** EVL cell fusions binned from sphere–30 % epiboly (4–4.66 hpf) and from 30–55 % epiboly (4.66–5.66 hpf) stages, plotted as mean ± s.e.m. in cell division inhibitor-treated (*n* = 6 embryos) or α-dynein antibody-injected embryos (*n* = 4 embryos); control embryos for the inhibitor and dynein antibody experiments were incubated in 1 % dimethylsulphoxide (*n* = 5 embryos) or injected with the antibody supernatant/ascites (*n* = 4 embryos), respectively; error bars, s.e.m. **(b)** Time course of an exemplary EVL cell fusion event (arrowheads) in a cell division inhibitor-treated embryo from sphere stage (*t* = 0 min) onwards. Cell membrane and spindle microtubules were outlined by GPI-RFP and Tau-GFP, respectively; scale bar, 20 μm. **(c)** Histogram showing the frequency distribution of the orientation of fusing EVL cell-cell junctions; *n* = 21 fusions, 6 embryos. **(d)** Average rates of apical cell-area increase for both fusing (squares, *n* = 23 fusions, 6 embryos) and non-fusing EVL cells (circles, *n* = 28 cell pairs, 6 embryos) in cell division inhibitor-treated embryos; error bars, s.e.m.; *P* = 0.53 (calculated by using cell pair/fusion numbers). **(e)** Exemplary images and average growth rate of fusing (arrowhead) and non-fusing (arrow) cell-cell junctions labeled with GPI-RFP in a cell division inhibitor-treated embryo; error bars, s.e.m.; *n* = number of cell pairs/fusions, 6 embryos; *P* < 0.0001; scale bar, 20 μm. Number of independent experiments: 19 **(a)**, 6 **(c,d)**.

**Figure-7: EVL cell fusions in cylindrically deformed embryos. (a)** Low magnification of an exemplary *Tg(actb2:GFP-utrCH)* cylindrical embryo at sphere (4

hpf; $t$ = 0 min) and 70 % epiboly stage (7.5 hpf; $t$ = 211 min). Arrowheads point to multinucleated EVL cells resulting from cell fusions; scale bar, 100 μm. **(b)** EVL cell fusions binned from sphere–30 % (4–4.66 hpf) and from 30–55 % (4.66–5.66 hpf) epiboly stage, plotted as mean ± s.e.m. in normal ($n$ = 6 embryos) and cylindrically deformed embryos ($n$ = 4 embryos). **(c)** Time course of an exemplary EVL cell fusion event (arrowheads) in a cylindrical embryo from sphere stage (4 hpf; $t$ = 0 min) onwards**.** Arrows point at cell divisions, of which one gave rise to a daughter cell subsequently undergoing fusion. Cell membrane and nuclei were marked by GPI-RFP and H2A-Cherry, respectively; scale bar, 20 μm. Number of independent experiments, 10 **(b)**.

Figure - 1 (Heisenberg)

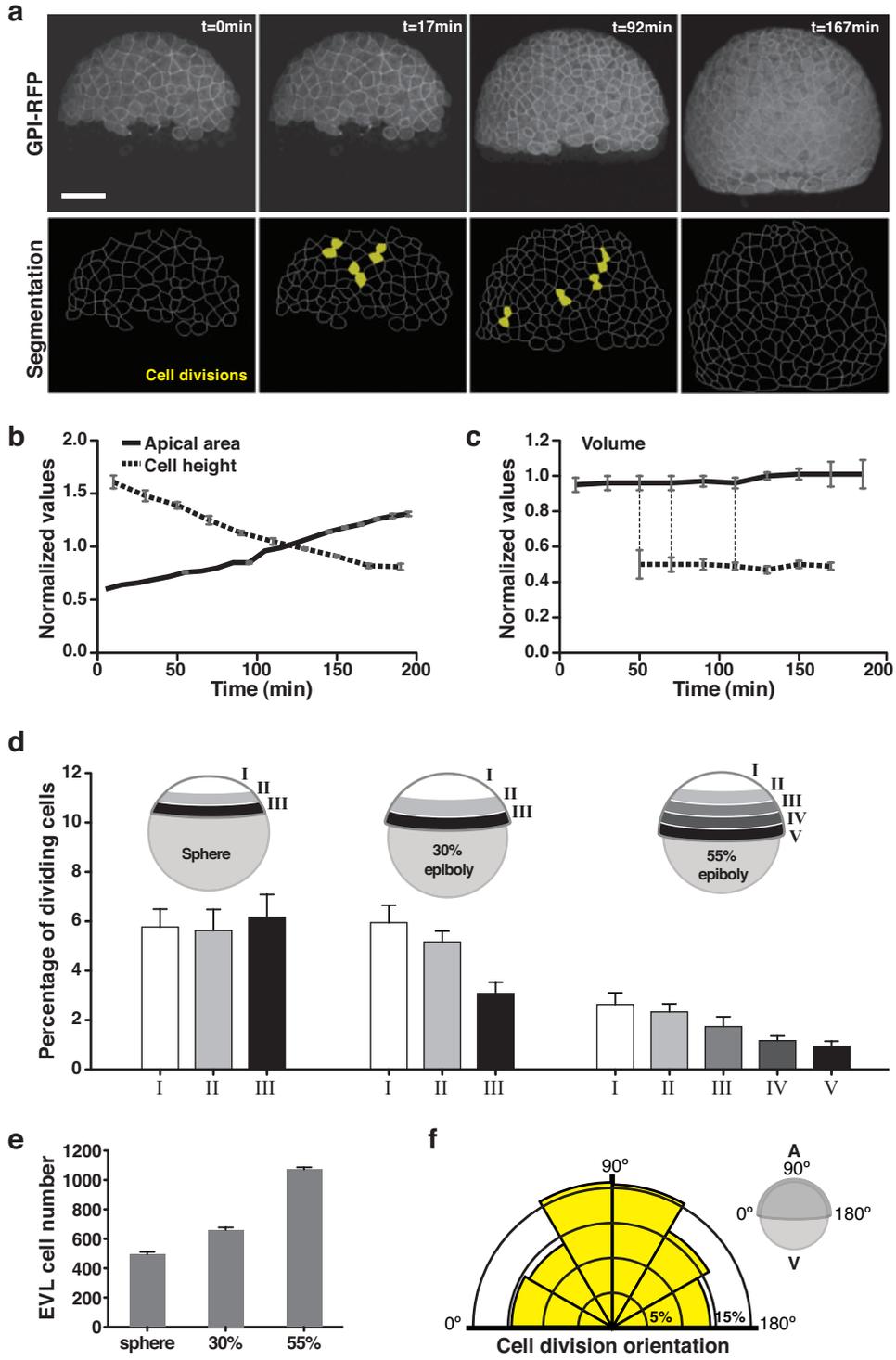

Figure - 2 (Heisenberg)

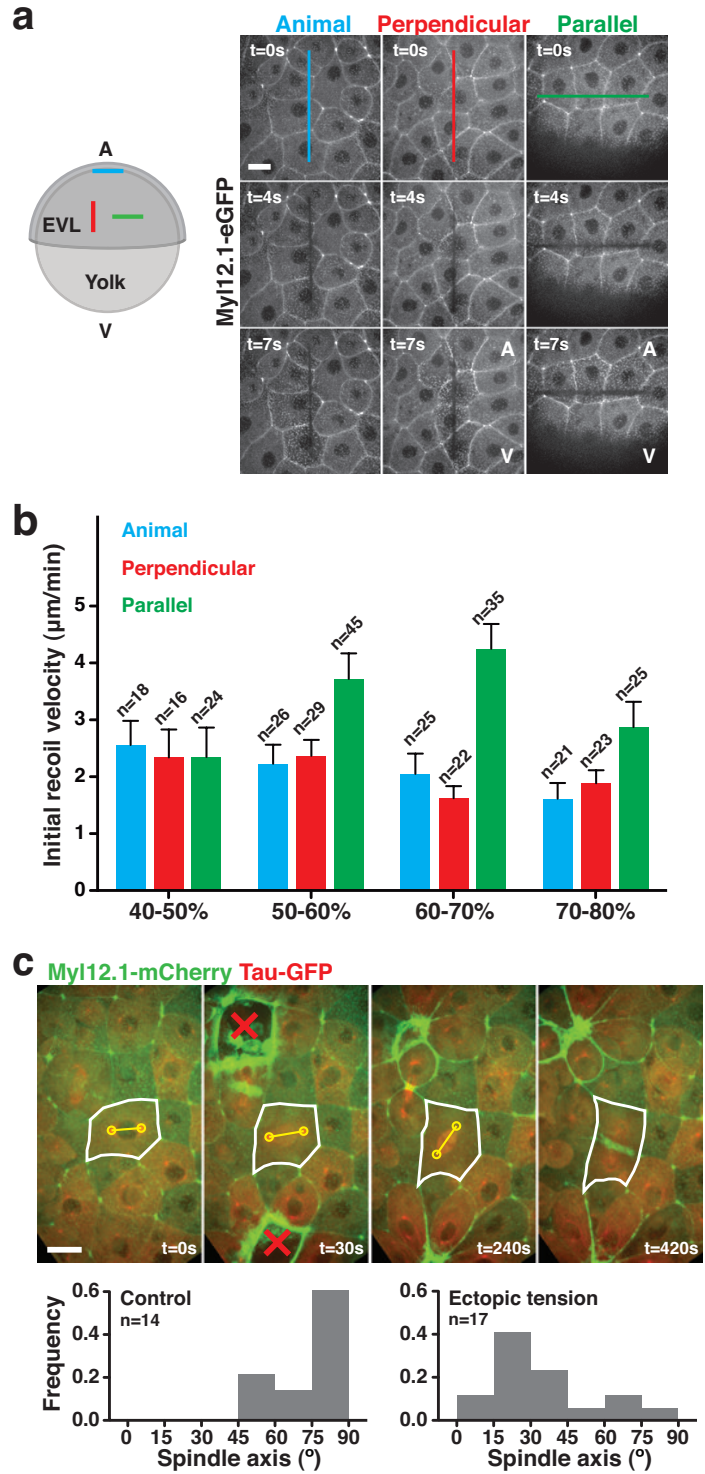

Figure - 3 (Heisenberg)

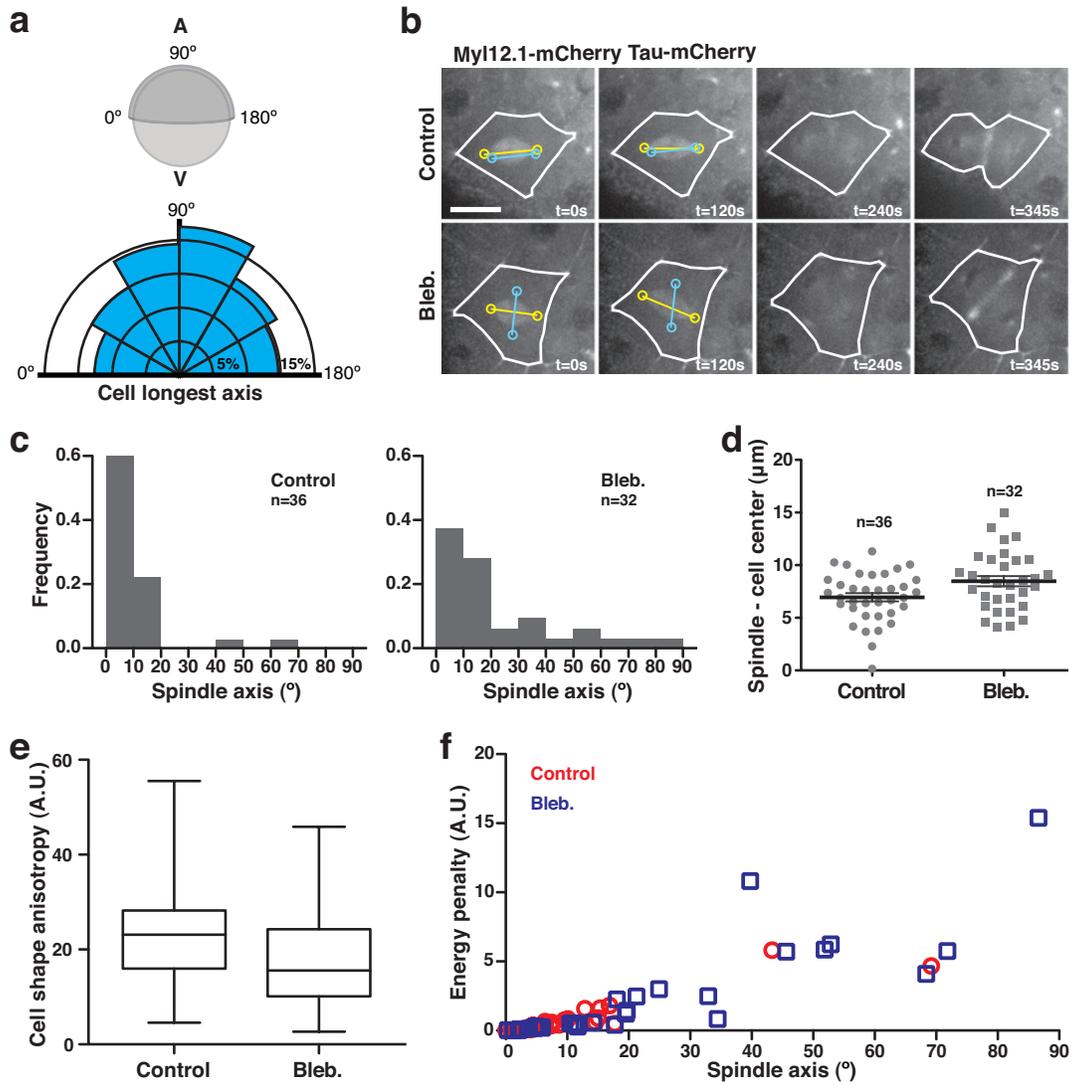

Figure - 4 (Heisenberg)

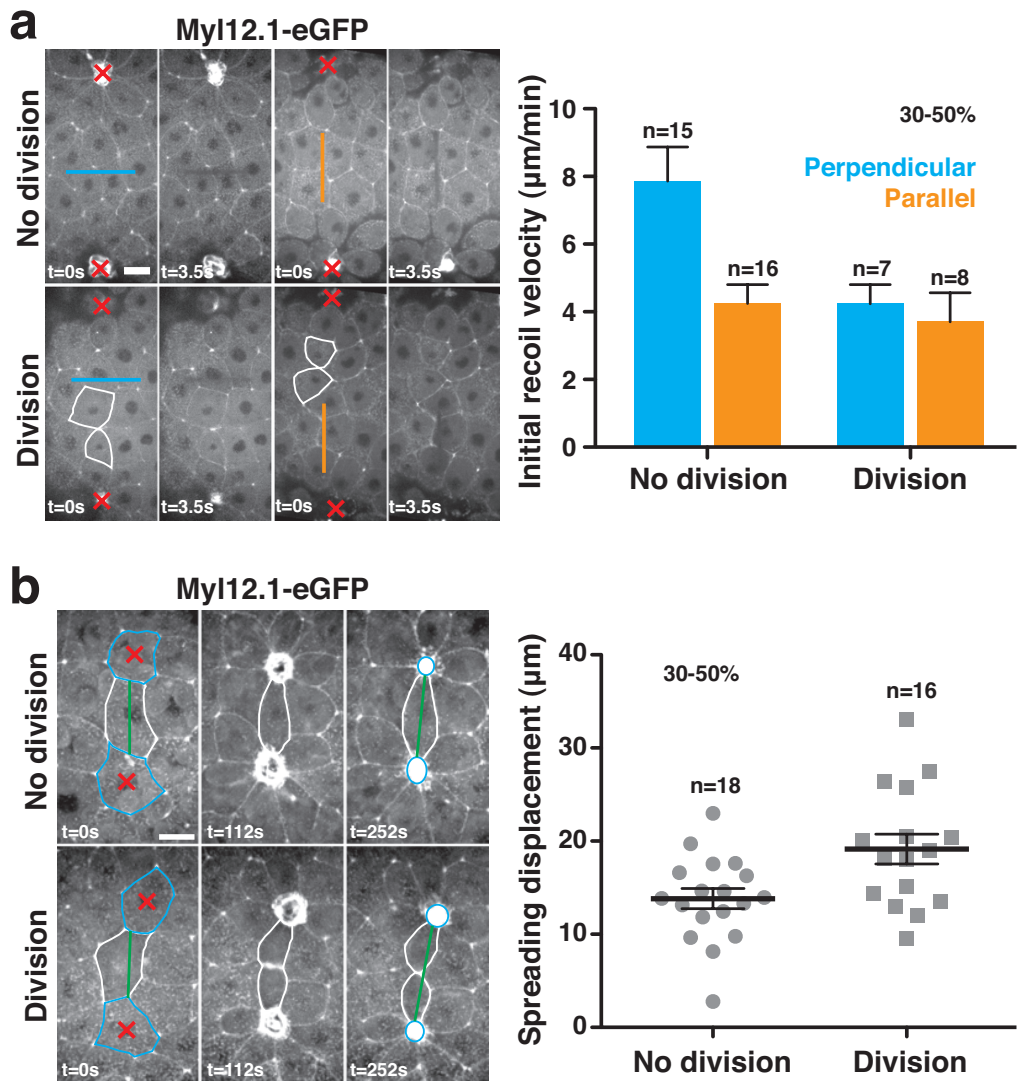

Figure - 5 (Heisenberg)

**a**

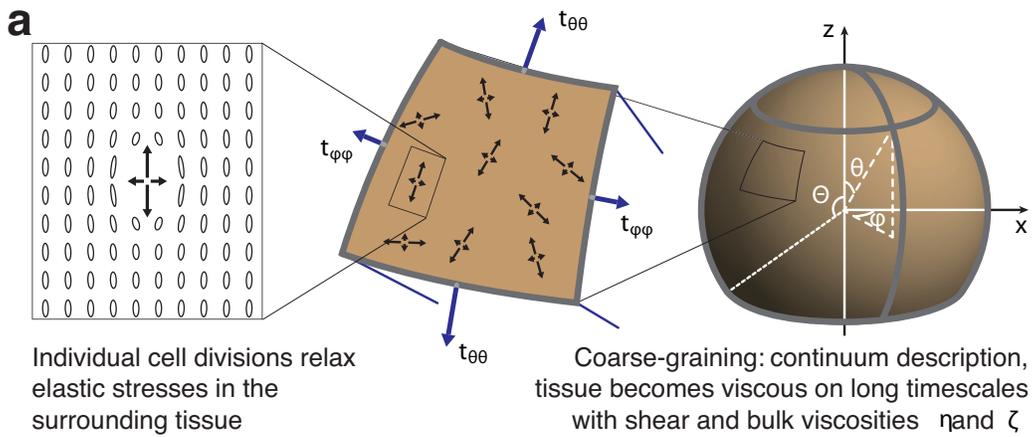

Individual cell divisions relax elastic stresses in the surrounding tissue

Coarse-graining: continuum description, tissue becomes viscous on long timescales with shear and bulk viscosities $\eta$ and $\zeta$

**b**

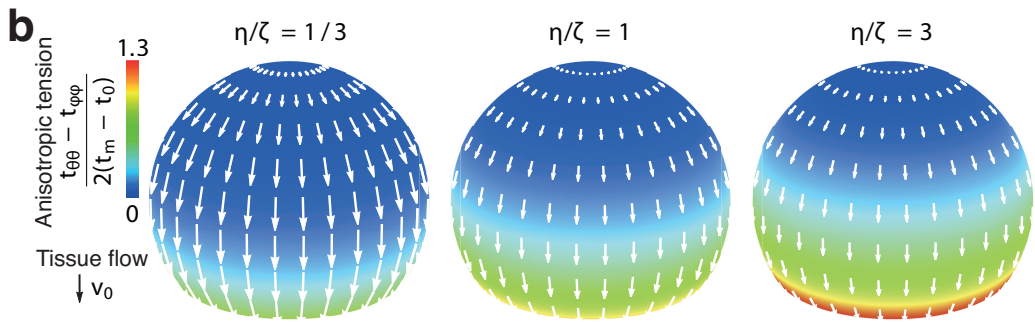

Tension-oriented cell divisions lower the shear viscosity: $\eta = \eta_0 \eta_{div} / (\eta_0 + \eta_{div})$

Model predictions: (1) Anisotropic tension distribution, stereotypic cell-division orientation
(2) Reduced tension-oriented cell division increases anisotropic tensions and reduces tissue flow

**c**

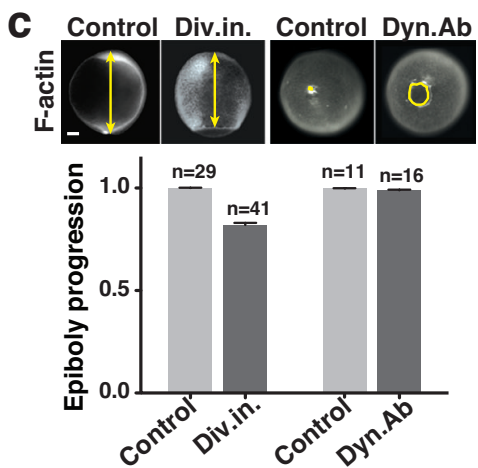

**d**

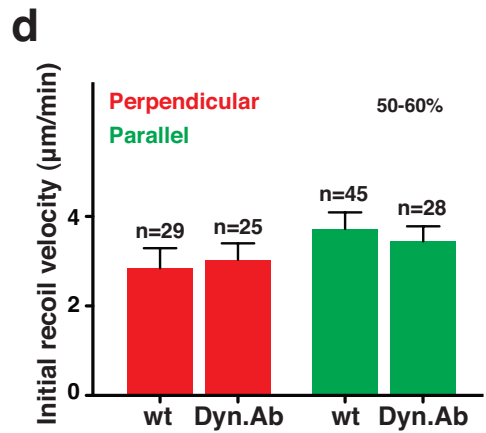

Figure - 6 (Heisenberg)

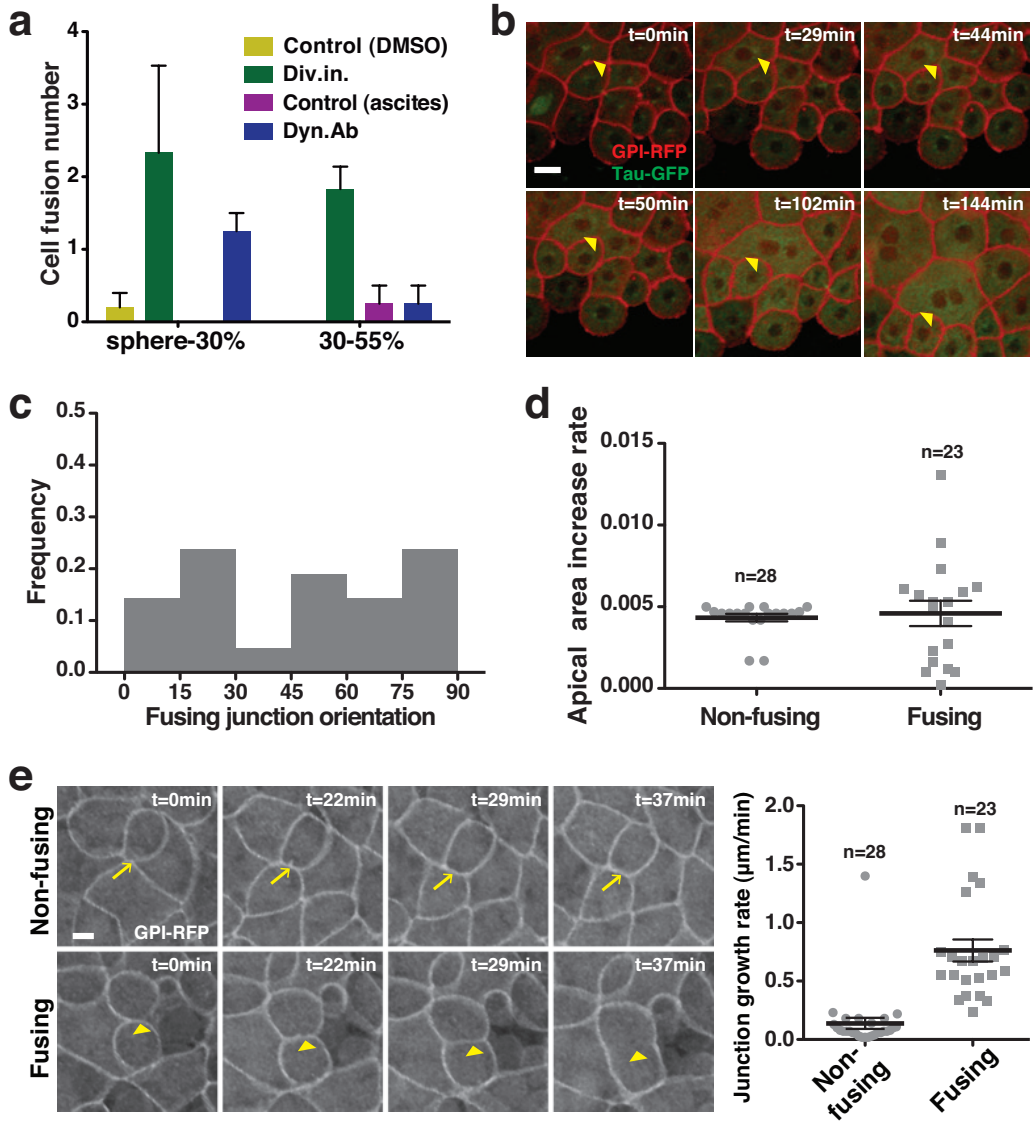

Figure - 7 (Heisenberg)

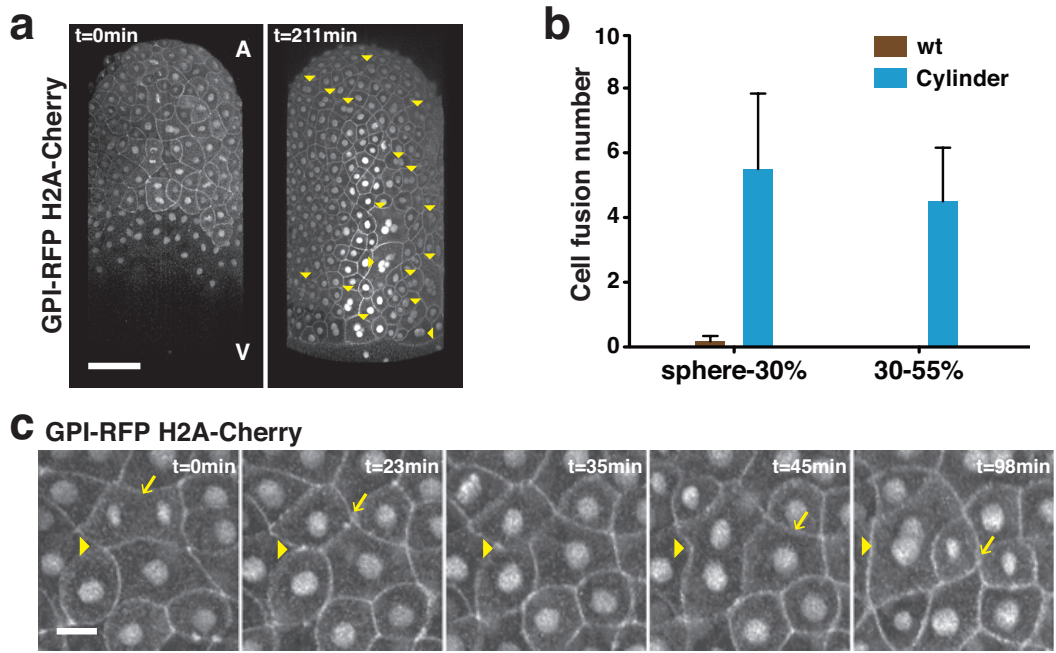